\documentclass[aps,apl,twocolumn,superscriptaddress,footinbib]{revtex4}
\usepackage{graphicx}
\usepackage{units}

\usepackage[version=3]{mhchem}
\usepackage[latin1]{inputenc}
\usepackage{setspace}
\newcommand{\superscript}[1]{\ensuremath{^{\textrm{#1}}}}
\newcommand{\subscript}[1]{\ensuremath{_{\textrm{#1}}}}

\begin{document}
\title{Direct growth of graphitic carbon on Si(111)}
\author{Pham Thanh Trung}
\author{Frédéric Joucken}
\email[Corresponding author: ]{fjoucken@fundp.ac.be}
\affiliation{University of Namur (FUNDP), Research Center in Physics of Matter and Radiation (PMR), 61 Rue de Bruxelles, 5000 Namur, Belgium}
\author{Jessica Campos-Delgado}
\author{Jean-Pierre Raskin}
\affiliation{Universit\'e Catholique de Louvain (UCL), Institute of Information and Communication Technologies,
Electronics and Applied Mathematics (ICTEAM), 4 Avenue Georges Lemaître, 1348 Louvain-la-Neuve, Belgium}
\author{Beno\^it Hackens}
\affiliation{Universit\'e Catholique de Louvain (UCL), IMCN/NAPS, 2 Chemin du Cyclotron, 1348 Louvain-la-Neuve, Belgium}
\author{Robert Sporken}
\affiliation{University of Namur (FUNDP), Research Center in Physics of Matter and Radiation (PMR), 61 Rue de Bruxelles, 5000 Namur, Belgium}

\begin{abstract}
Appropriate conditions for direct growth of graphitic films on Si(111) 7$\times$7 are investigated. The structural and electronic properties of the samples are studied by Auger Electron Spectroscopy (AES), X-ray Photoemission Spectroscopy (XPS), Low Energy Electron Diffraction (LEED), Raman spectroscopy and Scanning Tunneling Microscopy (STM). In particular, we present STM images of a carbon honeycomb lattice grown directly on Si(111). Our results demonstrate that the quality of graphene films formed depends not only on the substrate temperature but also on the carbon buffer layer at the interface.  This method might be very promising for graphene-based electronics and its integration into the silicon technology.
\end{abstract}

\maketitle
Nowadays, enormous efforts are devoted to grow and transfer graphene on various substrates using different methods such as mechanical exfoliation of highly oriented pyrolytic graphite (HOPG)\cite{Novoselov2004}, chemical vapor deposition (CVD) on metal substrate\cite{Wintterlin2009}, thermal decomposition of SiC in ultra-high vacuum (UHV)\cite{Hass2008}, etc. However, in order to integrate graphene into the current Si technology, it is highly desirable to grow graphene directly on silicon wafers. The formation of graphene on Si(111) has been a subject of previous research, not only due to its basic scientific interest but also to its technological significance\cite{Tang2011,Hackley2009,Ochedowski2012,Ritter2008}.
\\
The Si(111) 7$\times$7 surface has a complicated multi-layer reconstruction driven by the minimization of dangling bonds at the surface\cite{Quian1987}. It exhibits a six-fold symmetry in-plane so that it is expected to be an appropriate substrate for graphitic carbon growth.
However, due to the huge lattice mismatch between graphene sheets ($a_G=2.46\unit{~\AA}$) and Si(111) 7$\times$7 ($a_{Si_{7\times7}}=26.88\unit{~\AA}$), it is not easy to grow directly graphene at room temperature on Si(111) 7$\times$7. The Si(111) 7$\times$7 will reconstruct into 1$\times$1 at $\sim870\unit{~^{\circ}C}$\cite{Hibino1993}. At this temperature, the lattice mismatch between them is decreased to about 36\% and thus keeping the substrate at this temperature might be considered in order to grow graphene directly on Si(111) (although the mismatch is still important). However, Hackley et al.\cite{Hackley2009} showed that above $\sim700\unit{~^{\circ}C}$ the deposition of carbon leads to the formation of a SiC film instead of a graphitic film while below this temperature the carbon layer deposited is amorphous. They showed the importance of growing a carbon buffer layer at low temperature on the substrate prior to the growth of a graphitic film at higher temperature. Tang et al.\cite{Tang2011} reported different results: graphene films are formed when carbon is evaporated on the substrate above $\sim800\unit{~^{\circ}C}$ and amorphous carbon is found at lower temperature. Both studies used electron beam evaporators as carbon source.
\\
We present here results which demonstrate the crucial role of the buffer layer in order to grow graphitic films. We present furthermore STM images which establish unambugiously the graphitic nature of the films.
\\
The Si(111) ($\rho > 50\unit{~\Omega cm}$, n-type) samples are obtained by cycles of Ar\superscript{+} sputtering and annealing (up to 1000~°C) in UHV (pressure below $2.0\times10^{-10}\unit{~mbar}$) until a nice 7$\times$7 reconstruction is observed in LEED and STM. 
\\
The carbon source is a commercial e-beam evaporator from Tectra GmbH with a graphite rod of 99.997\% puriy from Goodfellow Cambridge Ltd.
\\
The samples are prepared {\textit{in situ} by evaporating carbon on the Si(111) surface at different temperatures (measured with an IR pyrometer). The carbon deposition rate is measured by a quartz crystal oscillator. The pressure in the chamber is kept below $1.0\times10^{-8}\unit{~mbar}$ during the evaporation.
\\
The carbon evaporation rate is held constant at $\sim 7\times10^{14}\unit{~atoms\cdot cm^{-2}\cdot min^{-1}}$ until the carbon flux is shut off. First, the samples are covered by a carbon layer with varying thickness at room temperature; this layer is called the buffer layer. Then, the substrate temperature is gradually increased (in about 4~min) to $820\unit{~^{\circ}C}$ and is maintained at this temperature for five minutes. The carbon flux is then shut off and the sample is nominally cooled down to $200\unit{~^{\circ}C}$ at $20\unit{~^{\circ}C\cdot min^{-1}}$, and then free-cooled to room temperature. Four different samples (\#1, \#2, \#3 and \#4) with different buffer layer thicknesses ($\sim 3.5\times10^{15}\unit{~atoms\cdot cm^{-2}}$, $\sim 5.2\times10^{15}\unit{~atoms\cdot cm^{-2}}$, $\sim 1.1\times10^{16}\unit{~atoms\cdot cm^{-2}}$ and $\sim 1.4\times10^{16}\unit{~atoms\cdot cm^{-2}}$, respectively) were analyzed. 
\\
LEED (Omicron), AES (Omicron) and STM (VP2 from Park Instrument) analyses were performed {\textit{in situ}} while Raman and X-ray Photoemission Spectroscopy (XPS) were performed after transportation in the atmosphere.
XPS analysis was made with a K-Alpha spectrometer from Thermo Scientific with a monochromated Al K$\alpha$ X-ray source (1486.6~eV) and a resolution of 0.1~eV. Raman scattering measurements were performed using a LabRam HR system with a 514~nm laser excitation source and an objective of 100$\times$. No outgassing was possible before performing the XPS and the Raman measurements.
After the {\textit{ex situ}} measurements, the samples were reintroduced in the UHV chamber and AES measurements (after outgassing the samples at $\sim350\unit{~^{\circ}C}$ for 20~min) gave similar results to those reported below. 
\\
The SiC and HOPG crystals used as references were analyzed in the same chambers after outgassing at $\sim600\unit{~^{\circ}C}$ for several hours (except for the XPS and Raman measurements). An oxide layer is still present on the SiC after such outgassing\cite{johansson1998} while the HOPG showed no oxygen contamination.
\\
Fig.~\ref{fig1}a and~\ref{fig1}b display the Auger spectra and their derivatives around the C\subscript{KLL} transition and compare them to the spectra of SiC and HOPG. 
\begin{figure}[h]
\begin{center}
\includegraphics[width=1\columnwidth]{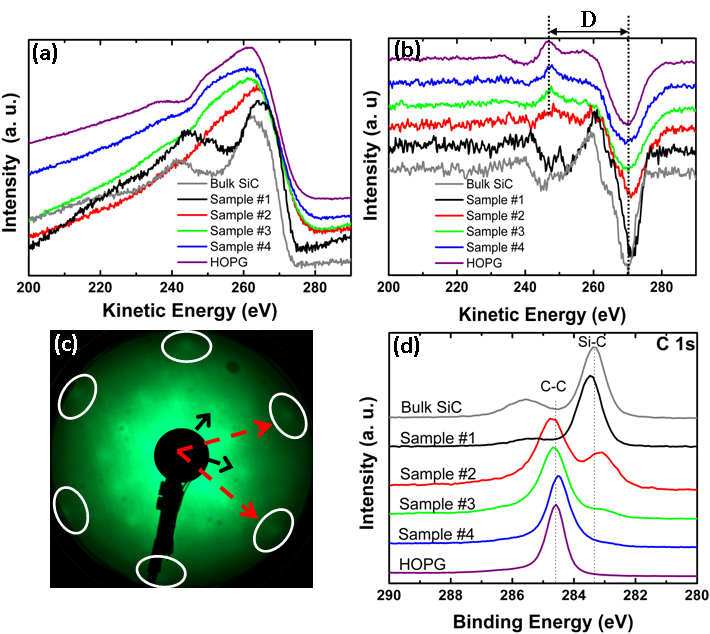}
\caption{a) AES spectra around the C\subscript{KLL} transition of the four samples as well as HOPG and SiC; b) The differentiated spectra; c) LEED pattern at 50.2~eV of sample \#1 showing spots corresponding to the SiC formation (lattice constant of $\sim3.1\unit{~\AA}$) d) C 1{\textit{s}} XPS spectra of samples \#1 to \#4 (and HOPG and SiC as references).}
\label{fig1}
\end{center}
\end{figure}
Clearly, one can see in Fig.~\ref{fig1}a that the shape of the curve of sample \#1 is similar to the one from the carbide while samples \#2, \#3 and \#4 are similar to the graphitic carbon signal (HOPG). The difference between the spectra appears more clearly on the differentiated spectra (Fig.~\ref{fig1}b). The energy differences D between the maximum and the minimum of the curve (illustrated in Fig.~\ref{fig1}b for the HOPG spectrum) are given in Table~\ref{tab1}. 
\begin{table}[t]
\centering
\caption{Values of D (cf. Fig.~\ref{fig1}b) for the four samples, SiC and HOPG (in eV).}
\begin{tabular}{c c c c c c}
 SiC & \#1  & \#2 &  \#3 &  \#4 & HOPG \\[1ex]
\hline \hline 
 11.0 & 11.0  & 22.0 & 22.6 & 22.6 &22.7
\end{tabular}
\label{tab1}
\end{table}
These differences are used to determine the ratio of {\textit{sp\superscript{2}}}-bonded carbon to {\textit{{\textit{sp\superscript{3}}}}}-bonded carbon in carbon compounds\cite{Mednikarov2005,Jackson1995}. We can conclude from those values that carbon atoms in sample \#1 are in the same state as in silicon carbide ({\textit{{\textit{sp\superscript{3}}}}} hybridization) while those in samples \#2, \#3 and \#4 are {\textit{sp\superscript{2}}}-bonded to other carbon atoms, as in HOPG.
\\
The SiC formation on sample \#1 is confirmed by its LEED pattern displayed in Fig.~\ref{fig1}c. There are six main diffraction spots (marked by circles and highlighted by red arrows) corresponding to a lattice constant of $3.1\unit{~\AA}$. This is consistent with 3{\textit{C}}-SiC(111) which is the SiC polytype expected to grow on Si(111) at these temperatures\cite{Matsunami1997}.
The black arrows point out diffraction spots that, although not well-resolved, could correspond to the $\sqrt{3}\times\sqrt{3}$ reconstruction which has been observed for this surface\cite{Suemitsu2010,Ouerghi2010}.
\\
The graphitic nature of the carbon film on samples \#2, \#3 and \#4 and the carbide nature of the film on sample \#1 are further confirmed by XPS data on C 1{\textit{s}} core level shown in Fig.~\ref{fig1}d. The spectrum of sample \#1 is very similar to the SiC spectrum (except for the component at 285.5~eV which corresponds to the native oxide found on SiC\cite{johansson1998}). The main peak of sample \#2 appears at 284.7~eV, corresponding to C-C bonding, while a weaker component corresponding to the SiC formation (which took place because of a too thin buffer layer) is seen at 283.2~eV. The spectrum of sample \#4 is practically identical to the one of HOPG, indicating a graphitic nature for the carbon film on this sample.
\\
Raman measurements were performed in the 1200-2800~cm\superscript{-1} range to investigate the vibrations related to C-C bonds in the samples. The spectra recorded are plotted in Fig.~\ref{fig2}, where baselines have been substracted.  Lorentzian fittings have been carried out in order to analyze quantitatively the data. 
A careful inspection of the data reveals that sample \#1 does not show the typical {\textit{sp\superscript{2}}} related signatures of C-C bonds, however a strong signal at $\sim 1450\unit{~cm^{-1}}$ appears (marked by *).  Such feature has been observed previously in amorphous SiC systems, showing its depletion as graphitization occurs in the systems\cite{Inoue1983,Calcagno2002}. This tendency is confirmed in our samples, as will be discussed below: graphitic bonds are present in the rest of the samples, accordingly, the intensities of the features at 1450~cm\superscript{-1} are less important (gray curves in Fig.~\ref{fig2}).
The $G$ band (at 1600~cm\superscript{-1}, green fitted bands in Fig.~\ref{fig2}) is present in samples \#2, \#3 and \#4 confirming the presence of graphitic bonds, in good agreement with our AES characterization.  The disorder-related features are also present in these samples ($D$ bands at 1350~cm\superscript{-1}, blue fitted bands and $D'$ band at 1620 cm-1, red fitted band). For decades the ratio of the intensities of the $D$ and $G$ bands ($I_D/I_G$ ratio) in carbonaceous materials has been calculated as a direct indication of the size of graphitic crystallites\cite{Lespade1984,Ferrari2000}. In this study, $I_D/I_G$ ratios of the integrated areas have been calculated as well as the related crystallite sizes ($L_a$) according to the formula proposed by Ferrari and Robertson\cite{Ferrari2000}:  $I_D/I_G=0.0055 L_a^2$ (supposing a regime in which our materials evolve from amorphous carbon to nanocrystalline graphite).  Analysis of the $I_D/I_G$ ratios of our materials show an increase in crystallite size: for sample \#2, $L_{a}=17\unit{~\AA}$; for sample \#3, $L_{a}=19\unit{~\AA}$ and for sample \#4, $L_{a}=22\unit{~\AA}$.  It is worth mentioning that for samples \#3 and \#4 the $G'$ band appears (orange fitted bands in Fig~\ref{fig2}), suggesting a higher degree of stacking order when compared to sample \#2.  Overall, the depletion of the 1450~cm\superscript{-1} feature and the presence of {\textit{sp\superscript{2}}} related features ($D$, $G$ and $G'$ bands) in our samples strongly confirm the growth of graphitic films on silicon substrates.
\begin{figure}[h]
\begin{center}
\includegraphics[width=1\columnwidth]{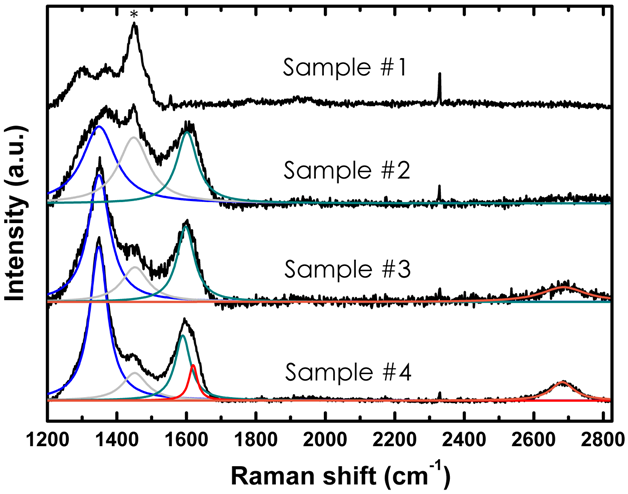}
\caption{Raman measurements of the studied samples, the different spectra have been vertically shifted to better illustrate the differences. The different peaks appearing in the spectra of samples \#2, \#3 and \#4 have been fitted to single Lorentzians.}
\label{fig2}
\end{center}
\end{figure}
\\
From the above analysis, we conclude that in order to grow graphitic carbon on Si(111) the minimum thickness of the buffer layer is $\sim 1.1\times10^{16}\unit{~atoms\cdot cm^{-2}}$ (sample \#2 marks the transition between SiC and graphitic carbon; sample \#3 being considered as graphitic).
\\
STM imaging strongly supports the previous conclusions. Fig.~\ref{fig3}a shows a large scale image of the sample~\#4. The steps of the Si(111) substrate are still clearly resolved but the root mean square roughness of the surface ($\sim1.2\unit{~\AA}$; between the substrate steps) is much higher than the one of the bare Si(111)7$\times$7 ($\sim0.3\unit{~\AA}$). Despite this roughness, we managed to achieve atomic resolution on samples \#2, \#3 and \#4 as shown in Fig.~\ref{fig3}b, c and d, respectively. 
\begin{figure}[h]
\begin{center}
\includegraphics[width=0.9\columnwidth]{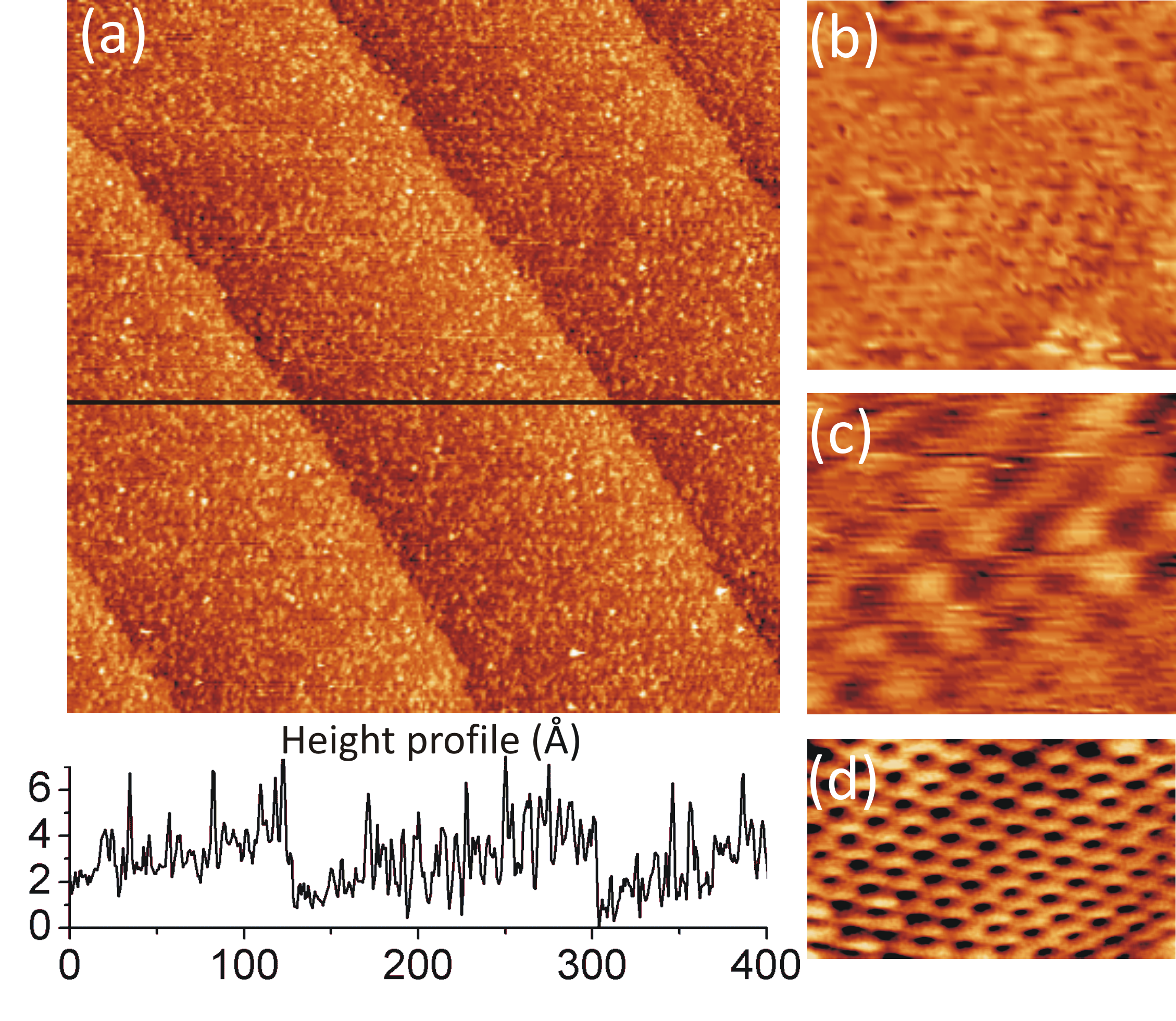}
\caption{STM images of samples \#2, \#3 and \#4. a) Large scale (400$\times$400~nm\superscript{2}) image of sample \#4 with a height profile ($V_{Sample}=+3\unit{~V}{\text{, }} I_{Tunnel}=0.35\unit{~nA}$); b) 2.5$\times$2.5~nm\superscript{2} image of sample \#2 ($V_{S}=-1\unit{~V}{\text{, }} I_{T}=6\unit{~nA}$; c) 1$\times$1~nm\superscript{2} image of sample \#3 ($V_{S}=-1.5\unit{~V}{\text{, }} I_{T}=4\unit{~nA}$; d) 2.5$\times$1.5~nm\superscript{2} image of sample \#4 ($V_{S}=-1\unit{~V}{\text{, }} I_{T}=4\unit{~nA}$) showing the honeycomb lattice of a graphene sheet.}
\label{fig3}
\end{center}
\end{figure}
Although the resolution of the images of samples \#2 and \#3 is not good, a triangular lattice is still visible. Height profile analysis reveals that the lattice constant is indeed $\sim2.5\unit{~\AA}$, as expected for graphitic surfaces. Those images present the triangular symmetry corresponding to the Bernal (ABA) stacking of the carbon layers\cite{Latil2007}. However, the image of sample \#4 (d) displays the honeycomb lattice of free-standing graphene. This can be explained by a rotational mismatch between the layer being scanned and the one underneath, restoring the symmetry between the two carbon atoms of the graphene unit cell\cite{Latil2007}. The observation of both the triangular and the honeycomb structure is similar to what has been reported already for HOPG\cite{Wang2006} and for epitaxial graphene on SiC(000\={1})\cite{Hass2008}. We must point out that the roughness of the surface as well as the small size of the crystallites (cf. Raman analysis) prevented us from reaching systematically the atomic resolution on the different samples. 
\\

In conclusion, we grew graphitic layers directly on Si(111) through the deposition of a buffer layer of amorphous carbon  at room temperature using electron beam evaporation. In particular, we obtained real space (STM) images of such films. However, the need for an amorphous buffer layer induces a roughness on the substrate that we believe limits the size of the graphitic nanocrystals that can possibly be obtained.
\\
P. T. T. and F. J. would like to thank Jacques Ghijsen for useful discussions and Etienne Gennart for technical support. B.H. is a FRS-FNRS research associate. This work was partially funded by ARC project no. 11/16-037.

\def\bibfont{\footnotesize}

\end{document}